\documentclass[aps,prl,reprint]{revtex4-1}
\usepackage{graphicx}
\usepackage{dcolumn}
\usepackage{bm}
\usepackage{amsmath}
\usepackage{siunitx}
\usepackage{xcolor}
\usepackage{eucal}

\begin{document}

\title{Measurement of the Low-temperature Loss Tangent of High-resistivity Silicon with a High Q-factor Superconducting Resonator} 
\author{M.~Checchin}
\email[]{checchin@fnal.gov}
\affiliation{Superconducting Quantum Materials and Systems Center, Fermi National Accelerator Laboratory, Batavia IL 60510, USA}
\author{D.~Frolov}
\affiliation{Superconducting Quantum Materials and Systems Center, Fermi National Accelerator Laboratory, Batavia IL 60510, USA}
\author{A.~Lunin}
\affiliation{Superconducting Quantum Materials and Systems Center, Fermi National Accelerator Laboratory, Batavia IL 60510, USA}
\author{A.~Grassellino}
\affiliation{Superconducting Quantum Materials and Systems Center, Fermi National Accelerator Laboratory, Batavia IL 60510, USA}
\author{A.~Romanenko}
\affiliation{Superconducting Quantum Materials and Systems Center, Fermi National Accelerator Laboratory, Batavia IL 60510, USA}

\date{\today}

\begin{abstract}
    In this letter, we present the direct loss tangent measurement of a high-resistivity intrinsic (100) silicon wafer in the temperature range from $\sim70$~mK to 1~K, approaching the quantum regime. The measurement was performed using a technique that takes advantage of a high quality factor superconducting niobium resonator and allows to directly measure the loss tangent of insulating materials with high level of accuracy and precision. We report silicon loss tangent values at the lowest temperature and for electric field amplitudes comparable to those found in planar transmon devices one order of magnitude larger than what was previously estimated. In addition, we discover a non-monotonic trend of the loss tangent as a function of temperature that we describe by means of a phenomenological model based on variable range hopping conduction between localized states around the Fermi energy. We also observe that the dissipation increases as a function of the electric field and that this behavior can be qualitatively described by the variable range hopping conduction mechanism as well. This study lays the foundations for a novel approach to investigate the loss mechanisms and accurately estimate the loss tangent in insulating materials in the quantum regime, leading to a better understanding of coherence in quantum devices. 
\end{abstract}


\maketitle 
Superconducting quantum circuits based on cavity-quantum electrodynamics  architecture~\cite{Wallraf_Nature_2004,Schoelkopf_Nature_2008, Devoret_Science_2013} represent a leading technology for constructing quantum processors and achieving quantum supremacy~\cite{Arute_etal_Nature_2019}. This technology utilizes the nano-fabrication processes developed by the semiconductor industry to manufacture integrated microwave circuits. Silicon (Si) is a vital material in integrated circuit technology and provides good trade-off between losses in the milli-Kelvin regime and process industrialization; therefore, it is the substrate of choice for superconducting quantum bits fabrication.

Because of the high dielectric constant ($\varepsilon_\text{r}'$=11.5 at low temperatures), a large fraction of the electromagnetic energy is stored in the silicon substrate~\cite{Wang_APL_2015}; hence, its contribution to the overall device energy loss can be substantial. Accurate knowledge of the Si loss tangent is thus pivotal for correctly estimating dissipation in quantum devices.

High-resistivity silicon ($\rho\geq5$~k$\Omega$-cm) at milli-Kelvin temperatures and microwave frequencies is generally assumed to introduce negligible dielectric losses compared to the native oxide and substrate-metalization intermixing layers found in typical superconducting quantum devices. However, no direct measurements of Si wafer loss tangent at milli-Kelvin temperatures have been presented thus far. Therefore, a reasonable large degree of uncertainty on the actual Si loss tangent value in the milli-Kelvin range still exists. Loss tangent data of Si at milli-Kelvin temperatures exist for silicon billets~\cite{Bourhill_PRApplied_2019}, which are not a representative sample of the wafers used to fabricate quantum bits, and for high-resistivity silicon at higher temperatures~\cite{Krupka_IEEEMicrowave_2006}.

In this letter, we present the direct measurement of the loss tangent of high-resistivity floating zone (FZ) silicon wafers in the temperature range from $\sim 70$~mK to $1$~K. We discover a non-monotonic trend of the loss tangent of silicon with temperature and an increasing trend of the dissipation with increasing electric field, that are inconsistent with known dissipation mechanisms encountered at these temperatures, such as two-level systems (TLS) absorption. Additionally, we demonstrate that the milli-Kelvin Si loss tangent is one order of magnitude larger than previously extimated~\cite{Woods_PRApplied_2019, Melville_APL_2020}.
\begin{figure*}[t]
\centering
\includegraphics[width=15cm]{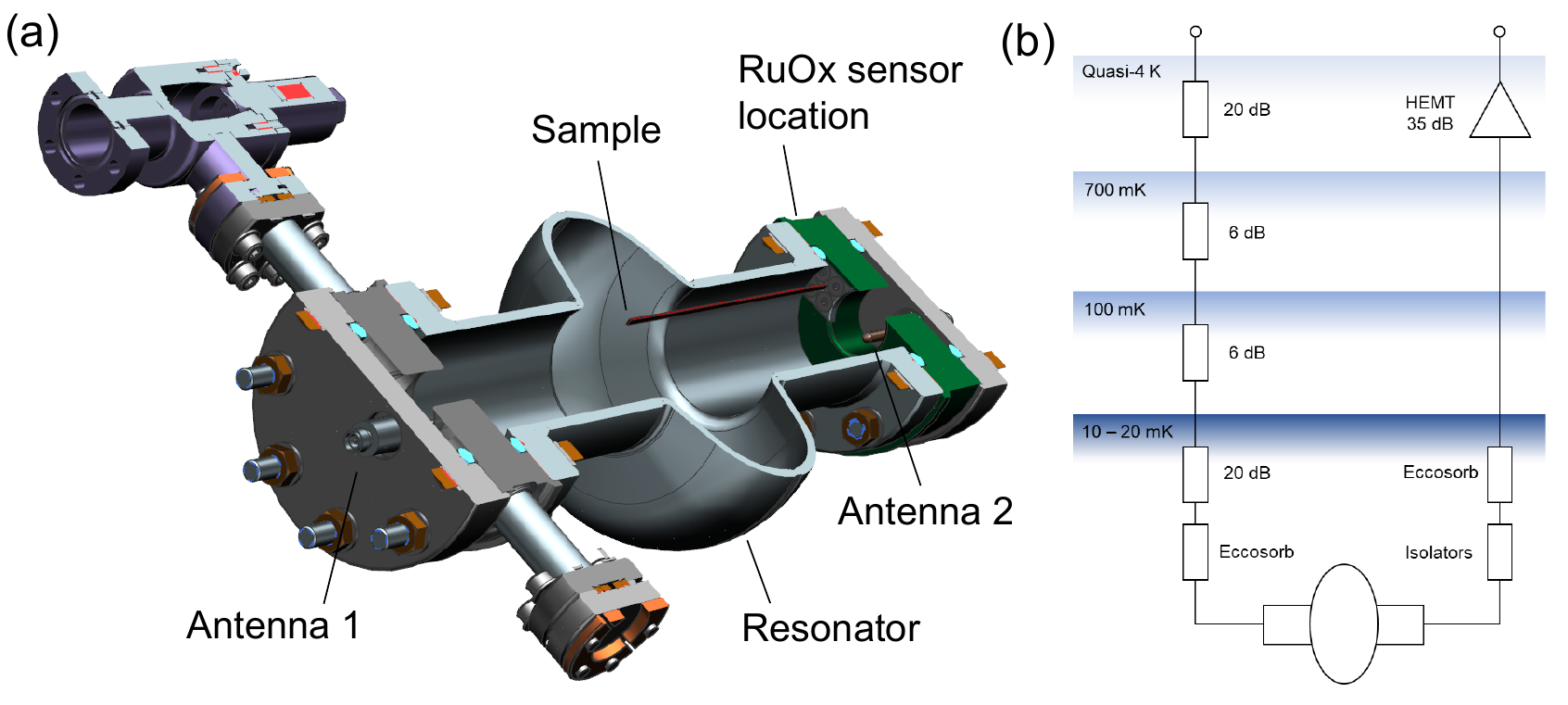}
\caption{Experimental set-up. (a) Three dimensional (3D) model of the experimental setup. (b) Schematic of the experiment in the dilution refrigerator. }
\label{fig.SetUp}
\end{figure*}

Measurements were performed using a high quality factor (Q-factor) elliptical superconducting niobium resonator hosting the Si sample in the high electric field region. The fundamental mode $\text{TM}_{010}$ resonating at 2.6~GHz was used to perform the sample characterization over the entire temperature range.

Elliptical superconducting resonators are typically adopted in particle accelerators at liquid He temperatures to accelerate relativistic charged particles because of their high efficiency (intrinsic Q-factor $Q_0\sim10^{11}$) in producing accelerating gradients on the order of tens of MV/m~\cite{Romanenko_APL_2014_2}. Nonetheless, they also allow for $Q_0\gtrsim 10^9$ at $10$~mK~\cite{Romanenko_PRApplied_2020}, making them an ideal tool for performing loss tangent measurements of insulators with a high degree of accuracy at milli-Kelvin temperatures.

Fig.~\ref{fig.SetUp}~(a) shows a 3D model of the experimental setup. The resonator was secured to the mixing chamber plate of a dilution refrigerator (DR). Temperature was recorded by two Ruthenium Oxide (RuOx) thermometers, one attached to the mixing chamber plate and one attached to the sample-holder transition flange, shown in green in Fig.~\ref{fig.SetUp}~(a)\textemdash the latter is referred to as sample thermometer. The input line had several stages of attenuation for a total of $\sim60$~dB, including the Eccosorb filters. The output line had isolators and Eccosorb filters connected immediately after the device output port and a 35~dB low-noise HEMT amplifier, which was thermalized to the quasi-4~K plate (typically stable at approximately 2.4~K). A simplified scheme of microwave connections in the DR is shown in Fig.~\ref{fig.SetUp}~(b). The total gain of the transmitted power line was measured to be approximately 62~dB, including a warm 40~dB amplifier (not shown in the schematics of Fig.~\ref{fig.SetUp}~(b)).

Intrinsic FZ Si(100) single-side polished wafers, with a thickness of 675~$\mu$m, were procured with room temperature resistivity of $10$~k$\Omega$-cm and diced to 10~cm long and 2~mm wide strips. After dicing, one sample was cleaned in an ultrasonic bath of isopropyl alcohol for 15~min, dried in ultra-pure nitrogen, assembled to the resonator, and subsequently pumped to a vacuum level of $p<10^{-5}$~Torr.

The loaded Q-factor ($Q_\text{L}$) was measured with a power ring-down technique. A steady-state electromagnetic field was established in the resonator and the transmitted power ($P_\text{t}$) free decay was recorded as a function of time after shutting off the power fed to the device. The peak electric field on the Si sample in the steady-state is measured to be approximately $E_{\text{Si}}=10$~V/m, corresponding to an average number of photons stored in the resonator in the order of $\langle n\rangle\sim5\cdot10^{9}$. This value was calculated by means of the proportionality factor $\kappa=E_{\text{Si}}/\sqrt{P_tQ_2}=E_{\text{Si}}/(\omega\sqrt{\hbar\langle n\rangle})=822$~V/(m-W$^{1/2}$) obtained by finite-element simulations by the high-frequency structure simulator (HFSS)
program, where $Q_2$ is the quality factor of the transmitted line antenna (Antenna 2 in Fig.~\ref{fig.SetUp}~(a)). To be noticed that independently on the number of photons in the resonator, the maximum electric field experienced by the Si sample in this experiment is comparable to that induced by plasma oscillations in the Si substrate of typical transmon devices, which is found to be up to 20~V/m in correspondence of the metalization edges~\cite{Lisenfeld_NPJ_2019}.

We define $Q_\text{L}=-10\omega/\left(\text{ln}10\cdot dP_\text{t}/dt\right)$, where $\omega$ denotes the angular frequency, and $dP_\text{t}/dt$ is the angular coefficient of the transmitted power free decay\textemdash with power measured in dBm. We then perform a linear regression to fit the decay data to extract $dP_\text{t}/dt$ selecting a window of points centered around 5~V/m. This strategy to measure $Q_\text{L}$ was implemented to increase the measurement accuracy and circumvent the typical S21 measurement approach that is limited by distortions of the resonant peak and appearance of side bands generated by microphonics~\cite{Padamsee_Book1} as a consequence of the high $Q_\text{L}$. Additional details regarding this measurement strategy are reported in Ref.~[\onlinecite{Romanenko_PRL_2017, Romanenko_PRApplied_2020}].

\begin{figure}[t]
\centering
\includegraphics[width=8cm]{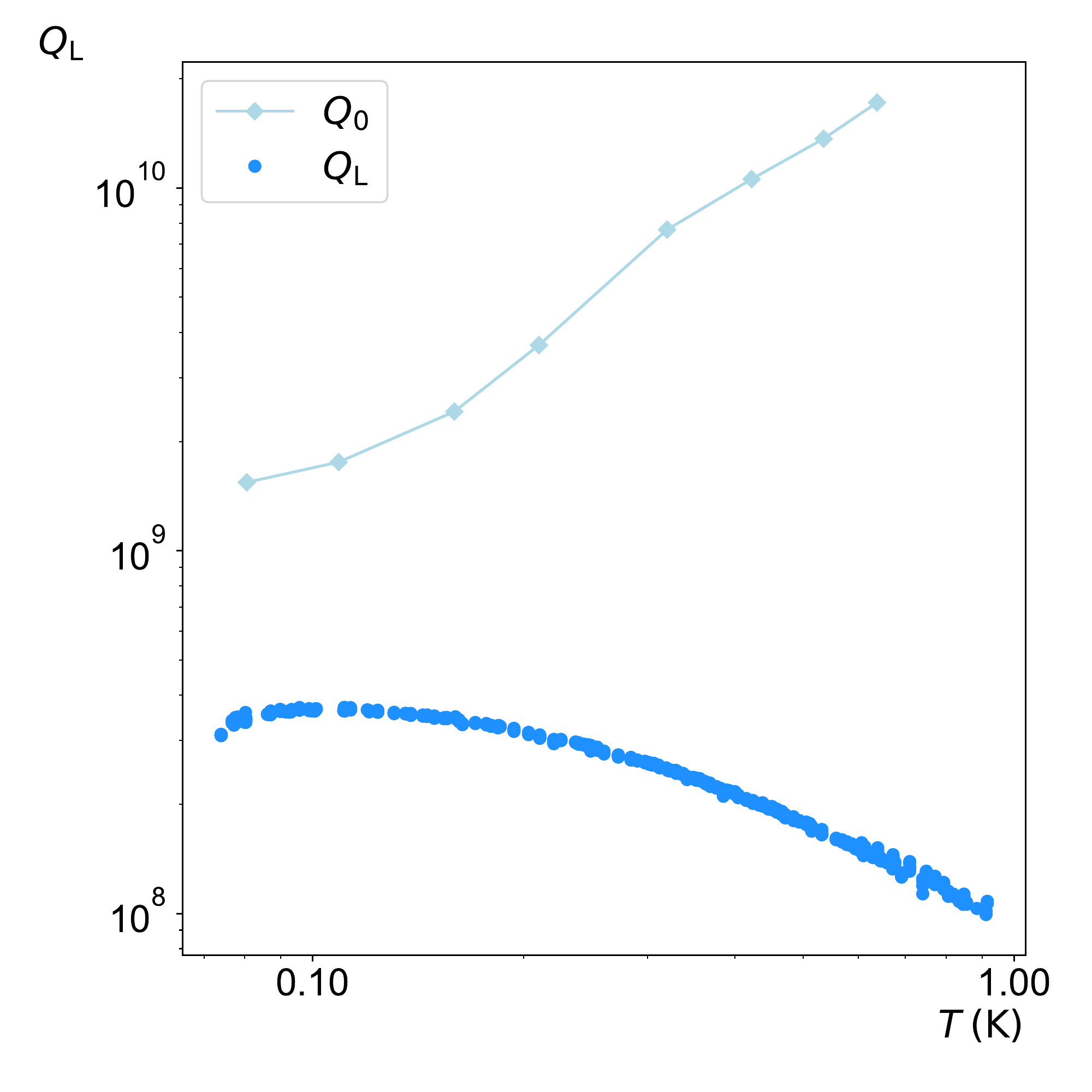}
\caption{The loaded quality factor as a function of the sample temperature is shown in blue. The resonator intrinsic Q-factor versus temperature at 2.6~GHz was reported in Ref.~[\onlinecite{Romanenko_PRApplied_2020}] and it is shown in light blue.}
\label{fig.QLvsT}
\end{figure}

In Fig.~\ref{fig.QLvsT}, we illustrate the loaded Q-factor measured as a function of temperature. The blue dots represent the loaded Q-factor ($Q_\text{L}$) data acquired in this study and plotted against the sample temperature. The light blue diamonds show the intrinsic Q-factor of the resonator alone, without the Si sample, as measured in Ref.~[\onlinecite{Romanenko_PRApplied_2020}]. At thermal equilibrium, the sample could not be cooled below $74$~mK due to a not ideal thermal path connecting the sample to the mixing chamber.

The loss tangent of the sample is calculated as follows:
\begin{equation}
    \dfrac{1}{Q_{\text{S}}}=\dfrac{p_\text{{Si}}}{Q_\text{{Si}}}+\dfrac{p_{\text{SiO}_2}}{Q_{\text{SiO}_2}}=\dfrac{1}{Q_\text{L}}-\dfrac{1}{Q_0}-\dfrac{1}{Q_1}-\dfrac{1}{Q_2}\text{,}
    \label{eq:Qsample1}
\end{equation}
where $Q_1=5.8\cdot10^9$ and $Q_2=6.5\cdot10^{11}$ represent the external Q-factors of the antennas (Antenna 1 and Antenna 2 in Fig.~\ref{fig.SetUp}~(a)) measured for the same setup in a liquid helium bath at $1.5$~K, as described in Ref.~[\onlinecite{Melnychuk_RevSciInstr_2014}], whereas $p_{\text{Si}}=9\cdot10^{-4}$ and $p_{\text{SiO}_2}=3\cdot10^{-9}$ denote the participation ratios of silicon and the native silicon oxide layer, defined as follows: $p_{\text{diel}}=\int_{V_{\text{diel}}}\varepsilon_{\text{diel}}|\mathbf{E}|^2dV_{\text{diel}}/\int_V\varepsilon_0|\mathbf{E}|^2dV$. Both values were calculated using the HFSS program.

It is important to highlight that the measured loaded Q-factor is dominated by the sample loss, whereas the other contributions are negligible. In fact, $Q_0$ is approximately one order of magnitude higher than $Q_\text{L}$ at the lowest temperature, while for temperature levels approaching 1~K, it is two orders of magnitude higher (see Fig.~\ref{fig.QLvsT}), and both $Q_1$ and $Q_2$ are at least one order of magnitude higher than $Q_\text{L}$. This implies that the measurement strategy implemented has a high level of accuracy and it is not affected by dissipation mechanisms extrinsic to the sample under study.

The loss tangent of SiO$_2$ at milli-Kelvin temperatures is experimentally known to be approximately $1/Q_{\text{SiO}_2}\simeq5\cdot10^{-3}$~\cite{Martinis_PRL_2005}. For the geometry under study, the participation ratio of the SiO$_2$ layer is calculated to be $p_{\text{SiO}_2}=3\cdot10^{-9}$. Therefore, SiO$_2$ contribution to $Q_\text{L}$ is $Q_{\text{SiO}_2}/p_{\text{SiO}_2}\sim10^{11}$, hence negligible compared to that of silicon. We can then define the silicon loss tangent as:
\begin{equation}
    \dfrac{1}{Q_{\text{Si}}}=\dfrac{1}{p_{\text{Si}}Q_{\text{S}}}\text{.}   
\end{equation}

\begin{figure}[b]
\centering
\includegraphics[width=8cm]{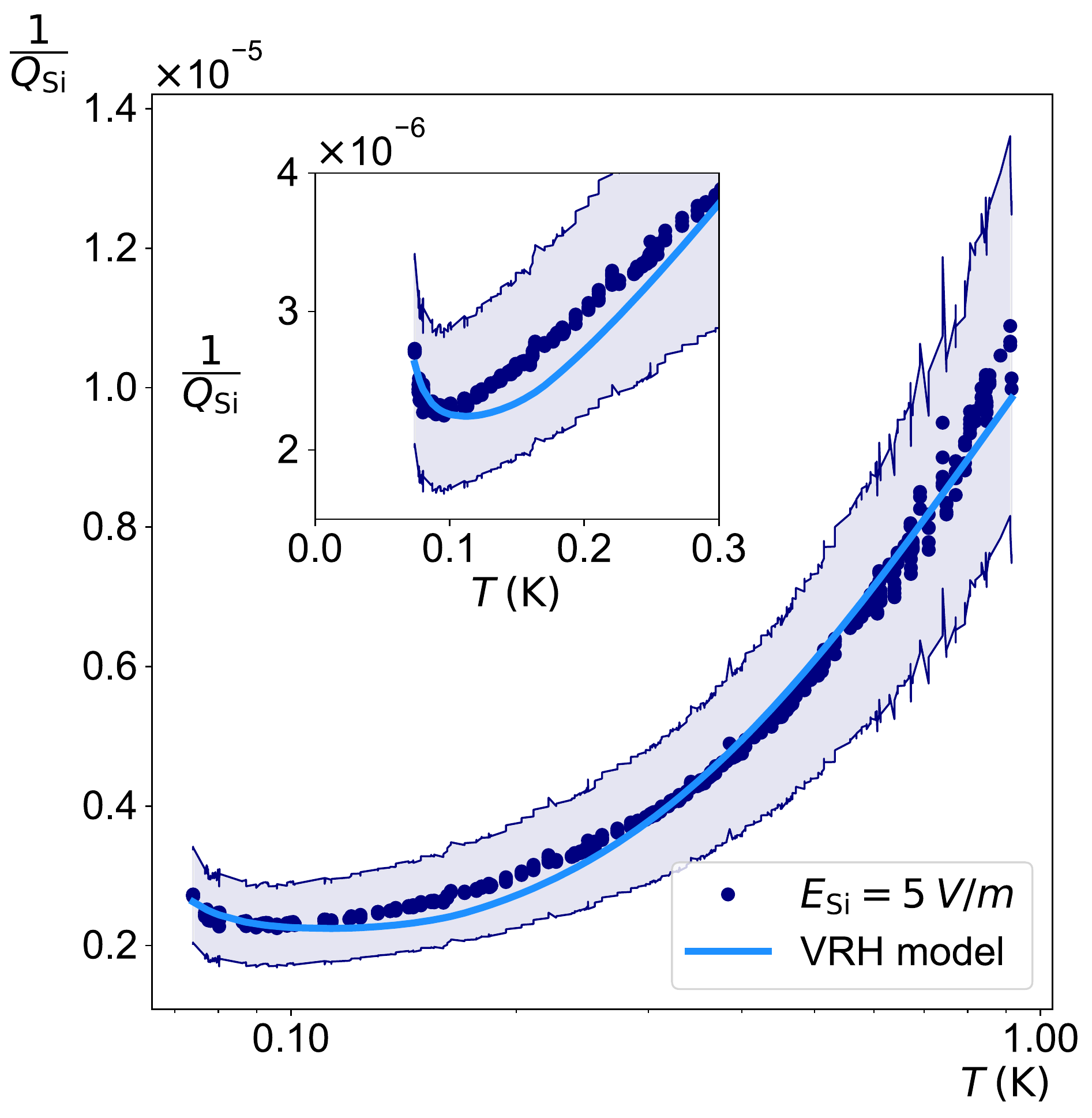}
\caption{Silicon loss tangent as a function of temperature. The blue solid line is a fit of the VRH model to the experimental data. The shaded area represent the experimental error of the measurement. The inset is a zoom of the data with temperature on a linear scale.}
\label{fig.lossTvsT}
\end{figure}

Fig.~\ref{fig.lossTvsT} shows the loss tangent of silicon as a function of the sample temperature. At the lowest temperature, 74~mK, the loss tangent is equal to $2.7\cdot10^{-6}$, which is in agreement with measurements performed on silicon billets in whispering gallery mode configuration~\cite{Bourhill_PRApplied_2019}. In contrast, our experiment recorded higher values compared to indirect estimations based on measurements and simulations of planar devices~\cite{Woods_PRApplied_2019, Melville_APL_2020}, which are in the order of $10^{-7}$, one order of magnitude lower than our experimental observations.

We observe a non-monotonic dependence of the silicon loss tangent on temperature, which does not resemble the two-level system (TLS) temperature dependence. The value of $1/Q_{\text{Si}}$ decreases with decreasing temperature, reaching a minimum at approximately 80~mK. Below 80~mK, the trend is opposite, and the loss tangent increases as the temperature decreases with a steeper slope. This trend is better appreciated in the inset of Fig.~\ref{fig.lossTvsT}, where the temperature is shown on a linear scale. The shaded areas show the experimental error associated with the measurement. This was calculated through error propagation, where the error in $Q_\text{L}$ was propagated from the square root of the variance of the angular coefficient that was calculated by the linear regression routine; the errors in $Q_0$, $Q_1$, and $Q_2$ were assumed to be $10$\% while that in $p_{\text{Si}}$ was estimated to be $25$\%, and it was calculated through HFSS simulations assuming $\pm 2$~mm sample misalignment and $11.5\leq\varepsilon_{\text{r}}'\leq11.9$.

\begin{figure}[t]
\centering
\includegraphics[width=8cm]{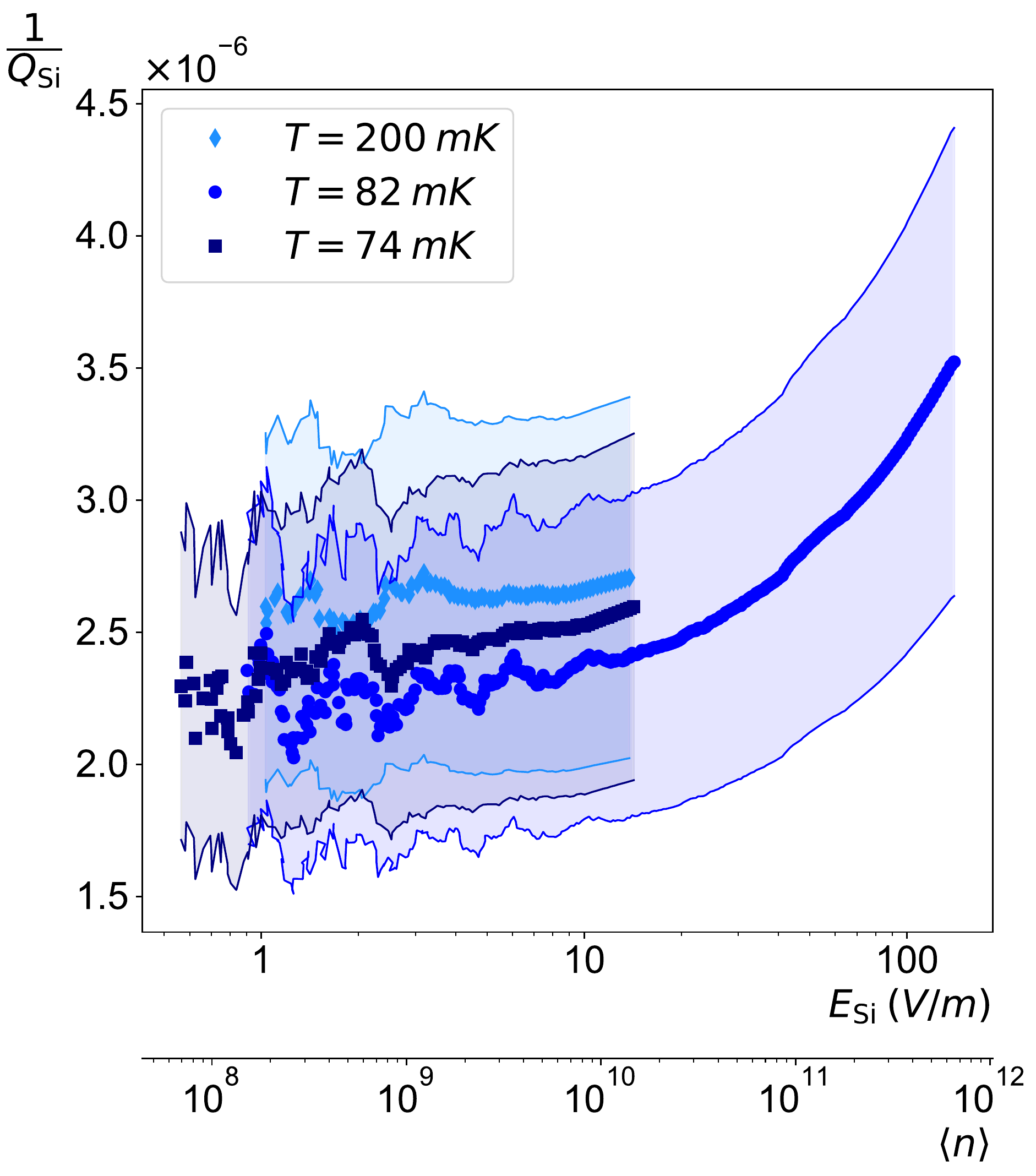}
\caption{Loss tangent as a function of the peak electric field in the sample measured at fixed temperature (74~mK, 82~mK, and 200~mK). The shaded areas represent the measurement error estimated as discussed in the text above. The second $x$ axis reports the number of photons in the resonator.}
\label{fig.lossTvsE}
\end{figure}

To further shed light into the dissipation mechanism in place as a function of temperature, we calculate the electric field dependence of the loss tangent at constant temperature. We select three temperatures above, equal, and below the minimum of the loss tangent as a function of T, nominally 74~mK, 82~mK, and 200~mK. And we calculate the silicon loss tangent by means of Eq.~\ref{eq:Qsample1}, where the loaded Q-factor is inversely proportional to the local slope of the transmitted power ring-down data ($Q_\text{L}=-10\omega/\left(\text{ln}10\cdot dP_\text{t}/dt\right)$). In Fig.~\ref{fig.lossTvsE}, we report the parametric plot of Si loss tangent versus peak electric field $E_\text{Si}$. Interestingly, the electric field dependence of the loss tangent is also not in agreement with TLS-driven dissipation. $1/Q_{\text{Si}}$ decreases with decreasing field, and saturates to a constant value for $E_{\text{Si}}\lesssim10$~V/m. The observed trend is independent on temperature\textemdash for temperatures above, equal, and below the loss tangent minimum the field dependence is unchanged. This finding points against the possibility that the upward trend of $1/Q_{\text{Si}}$ for temperatures below 80~mK is due to TLS-type of losses.

\begin{figure}[b]
\centering
\includegraphics[width=8cm]{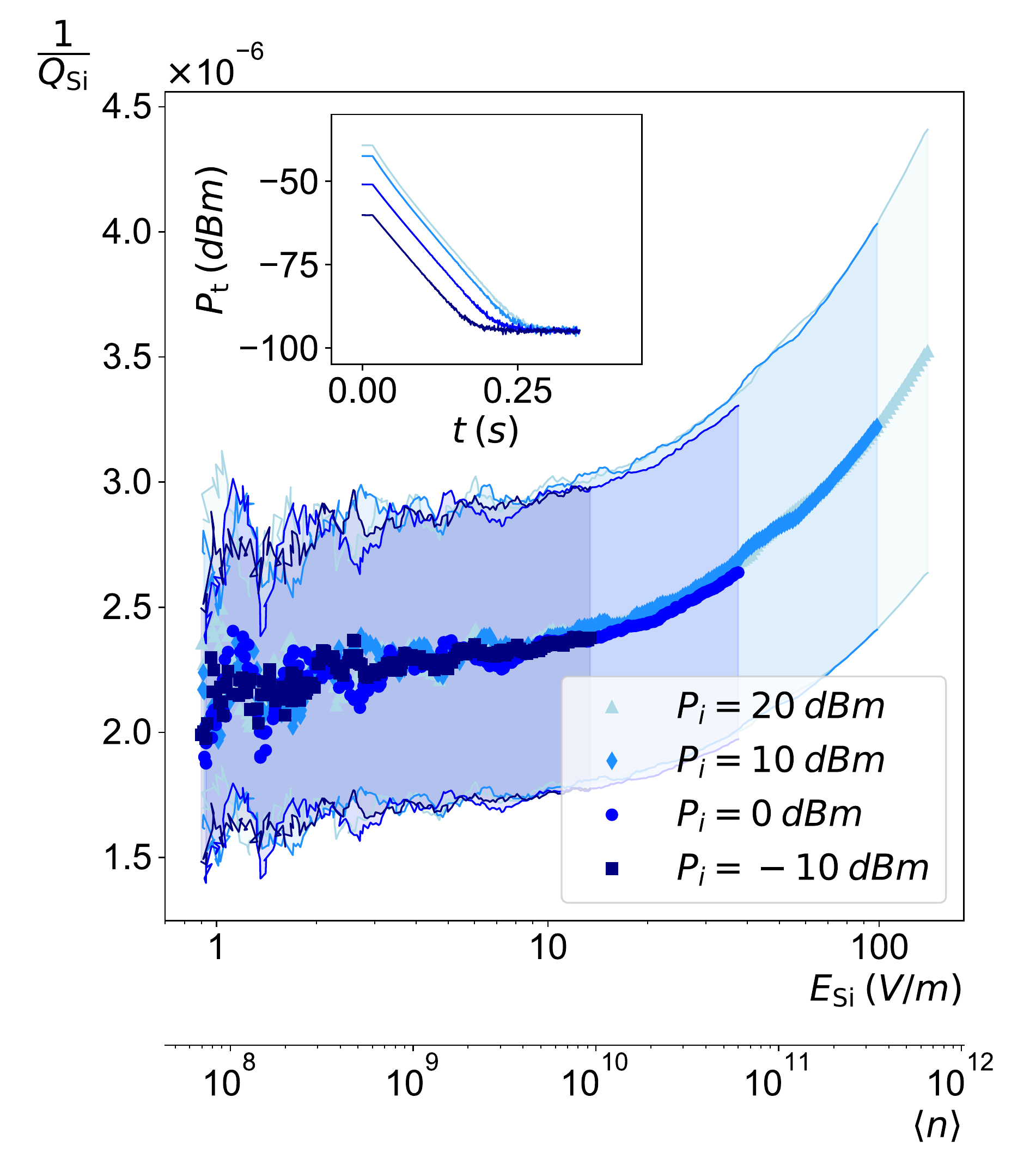}
\caption{Loss tangent as a function of the peak electric field on the sample measured at 82~mK for different steady-state $P_\text{i}$ values. The shaded areas represent the measurement error estimated as discussed in the text above. In the inset we report the power ring down data as a function of time. The color palette of the data reported matches the colors of the main plot.  The second $x$ axis reports the number of photons in the resonator.}
\label{fig.lossTvsE_pi}
\end{figure}

We exclude the possibility that the observed trend as a function of the electric field is due to an increase in temperature of the sample because of thermal feedback. To prove this, we performed a series of free decays experiments at fixed temperature (82~mK) with increasing input power ($P_\text{i}$) values, the experimental data is shown in Fig.~\ref{fig.lossTvsE_pi}. The thermalization time of the sample thermometer was observed to be roughly $\tau_\text{T}\sim2$~min, while the loaded decay time of the resonator plus sample was measured to be in the order of $\tau_\text{L}\sim0.1$~s. Since $\tau_\text{L}\ll\tau_\text{T}$, if any thermal feedback effect was at work then differences in the field dependence as a function of the steady-state $P_\text{i}$ would be expected. On the contrary, the electric field dependence of the loss tangent is independent on the steady-state condition, meaning that no feedback effect is playing a role in the measurements performed.

As discussed above, the experimental findings reported in this paper give insights into an unexplored dissipation mechanism of intrinsic Si at milli-Kelvin temperatures. Based on our experimental findings, the source of dissipation should be sought in mechanisms that are not TLS-related. Rather, we expect that conduction losses could be the cause of the observed temperature and field dependencies. Including conduction losses in the picture, the dielectric constant takes the form: 
\begin{equation}
    \varepsilon=\varepsilon_{\text{r}}' -i\left( \varepsilon_{\text{r}}'' +\dfrac{\sigma}{\omega\varepsilon_0}\right)\text{,}
\end{equation}
where $\varepsilon_{\text{r}}'$ and $\varepsilon_{\text{r}}''$ represent the real and imaginary parts of the complex dielectric permittivity, whereas $\sigma$ denotes the low temperature conductivity. The loss tangent is thereafter expressed as follows:
\begin{equation}
    \dfrac{1}{Q_{\text{Si}}}=\dfrac{1}{Q_{\text{TLS}}}+\dfrac{\sigma}{\omega\varepsilon_{0}\varepsilon_{\text{r}}'}\text{,}
    \label{eq:lossT}
\end{equation}
where the first term $1/Q_{\text{TLS}}=\varepsilon_{\text{r}}''/\varepsilon_{\text{r}}'$ denotes the dielectric loss tangent dominated by dielectric dissipation (TLS), whereas the second term describes the conduction losses. Due to the rationale discussed above, we can deem the TLS contribution of the Si oxide negligible.

We identify the dependence of the loss tangent with temperature with the second term of Eq.~\ref{eq:lossT}, and particularly with the occurrence of an electron hopping mechanism, likely variable range hopping (VRH)~\cite{Mott_JNonCrystSolids_1968,Apsley1_PhilMag_1974,Apsley2_PhilMag_1974}. VRH conduction may take place at low temperatures where localized states close to the Fermi level within $\sim\kappa_\text{B}T$ can contribute to the overall conduction. Electrons subjected to an electric field $\mathbf{E}$ hop between states separated by the shorter four-dimensional distance\textemdash three spatial coordinates and one energy coordinate, the so-called hopping range ($\mathcal{R}$), defined as:
\begin{equation}
\begin{split}
    &\mathcal{R}=2\alpha R+\dfrac{W+e\mathbf{R}\cdot\mathbf{E}}{\kappa_\text{B}T}\\
    &W=\dfrac{3^4}{4^4\pi g(\epsilon_\text{F}) R^3}\text{,}
\end{split}
    \label{eq:range}
\end{equation}
where $\alpha$, $W$, and $R$ are the wave-function localization parameter, average energy difference between the states~\cite{Mott_JNonCrystSolids_1968,Mott_Book}, with $g(\epsilon_\text{F})$ the density of localized states within $\sim\kappa_\text{B}T$ around the Fermi energy $\epsilon_\text{F}$, and the spatial distance between the states involved, respectively. The VRH conductivity can then be defined as proportional to the difference between the hopping probability with and against the field $\mathbf{E}$, $\langle P^+\rangle$ and $\langle P^-\rangle$:
\begin{equation}
    \sigma_\text{h}\sim\left|\langle P\rangle^+-\langle P\rangle^-\right|=\left|e^{-\text{min}\,\mathcal{R}^+}-e^{-\text{min}\,\mathcal{R}^-}\right|\text{.}
\end{equation}

In the zero field approximation, the distance $R$ that minimizes $\mathcal{R}$ in Eq.~\ref{eq:range} always increases as the temperature lowers and conduction becomes dominated by hops between levels that are closer in energy even if far apart, since the tunneling contribution (first right-hand side term in the first formula of Eq.~\ref{eq:range}) becomes exponentially suppressed compared to the energy activation term (second right-hand side term in the first formula of Eq.~\ref{eq:range}). Under this approximation, the conductivity is expected to decrease with $\sim\text{exp}\left(-T^{-1/4}\right)$, and for $T\rightarrow0$ it exponentially approaches zero~\cite{Mott_JNonCrystSolids_1968,Mott_Book}.

On the contrary, when $E>0$ then $R$ increases with different rates whether hops are happening with or against the electric field, with $R^-$ growth rate being the largest. In return, due to the $1/(R^-)^3$ dependence of $W^-$ and the increasing weight of $-eR^-E$ that lowers the hopping activation energy, the VRH process is governed by $\mathcal{R}^-$ since growing slower than $\mathcal{R}^+$. As in the $E=0$ case, $\sigma_\text{h}$ decreases with decreasing $T$. However, by further lowering the temperature, the condition $W^-<eR^-E$ is eventually met and electron hopping against $\mathbf{E}$ is not anymore limited by the activation energy. Consequently, the $\mathcal{R}^-$ range  reverts its trend and decreases with decreasing temperature, resulting in a sudden increase of $\sigma_\text{h}$ (increase of $\langle P^-\rangle$, decrease of $\langle P^+\rangle$) and forming a minimum that eventually translates in a minimum of $1/Q_\text{Si}$ versus $T$.

We fit the experimental data as a function of temperature by means of a phenomenological model based on the VRH mechanism described above. We rewrite Eq.~\ref{eq:lossT} assuming two conduction loss channels, specifically VRH (defined by $\sigma_\text{h}$) and a residual term (defined by $\sigma_0$) that allows for a better fit outcome. Eq.~\ref{eq:lossT} is rewritten as:
\begin{equation}
    \dfrac{1}{Q_{\text{Si}}}=\dfrac{\sigma_\text{h}+\sigma_0}{\omega\varepsilon_{0}\varepsilon_{\text{r}}'}\text{,}
\end{equation}
where the hopping conductivity is calculated as:
\begin{equation}
    \sigma_\text{h}=\dfrac{\left|J^+-J^-\right|}{E}\text{,}
\end{equation}
with $J=2e\gamma g(\epsilon_\text{F})\kappa_\text{B}TR\langle P\rangle$ and $\gamma$ the hopping attempt frequency~\cite{Mott_JNonCrystSolids_1968,Mott_Book}. 

The least square regression routine is ran by numerically calculating the values $R^+$ and $R^-$ that minimize $\mathcal{R}^+$ and $\mathcal{R}^-$ respectively, and by keeping $\alpha$, $\gamma$, $g(\epsilon_\text{F})$, and $\sigma_0$ as free parameters. The fit is reported in Fig.~\ref{fig.lossTvsT}. As shown, the model describes the data exhaustively within the experimental error, returning the following values: $\alpha^{-1}=1.05~\mu\text{m}$, $\gamma=11.4$~THz, $g(\epsilon_\text{F})=1.33\cdot10^{13}~\text{eV}^{-1}\text{cm}^{-3}$, and $\sigma_0=0.52~\mu$S/m. We interpret the $\sigma_0$ contribution as the manifestation of excitation of free-carriers in conduction band likely generated by cosmic rays absorption\textemdash due to the large dimension of the sample, a non-negligible fraction of excitations in conduction band are expected. Due to the lack of literature on VRH in high-resistivity intrinsic silicon, we compare the values returned by the fitting routine with those of amorphous silicon. The $\gamma$ value obtained falls within the range $\gamma\simeq(3-15)$~THz reported in Ref.~[\onlinecite{Pfeilsticker_ZPhys_1978}], while $g(\epsilon_\text{F})$ and $\alpha$ are smaller than what reported for amorphous silicon where $g(\epsilon_\text{F})\simeq(10^{16}-10^{23})~\text{eV}^{-1}\text{cm}^{-3}$ and $\alpha^{-1}\simeq(0.1-10)~\text{nm}$~\cite{Pfeilsticker_ZPhys_1978,Pichon_APL_2011}. These findings imply that i) as expected the number of available states in high-resistivity intrinsic silicon are less and separated by a larger distance compared to amorphous silicon, and ii) that the states involved in the hopping conduction mechanism are not well localized.

The VRH mechanism can qualitatively describe the experimental data as a function of the electric field as well, and for low $E$ values the conductivity is expected to increase with field as $\sigma_\text{h}\sim\text{sinh}(eRE/\kappa_\text{B}T)$~\cite{Mott_Book}. However, the fitting routine does not interpolate the data exhaustively, suggesting that a generalized first principles theoretical model is needed to describe accurately the field dependence observed. Nevertheless, the model here presented provides a first glimpse into the suspected loss mechanism at play at milli-Kelvin temperatures.

In conclusion, we described an accurate method to directly measure the loss tangent of insulating materials in wafer form using a high Q-factor superconducting resonator. We reported the direct measurement of the loss tangent of a high-resistivity silicon wafer in the temperature range from $\sim70$~mK  to 1~K. Furthermore, we showed that the loss tangent of high-resistivity Si in the milli-Kelvin range is one order of magnitude higher than that previously indirectly estimated from measurements and simulations of planar devices~\cite{Woods_PRApplied_2019,Melville_APL_2020}, although it is in agreement with values obtained from whispering gallery measurements of Si billets~\cite{Bourhill_PRApplied_2019}. In addition, we discovered a non-monotonic behavior of the loss tangent dependence on temperature with a minimum at approximately 80~mK, that we interpret as the occurrence of the sign change in the against-field hopping range ($\mathcal{R}^-$) slope as a function of temperature. We also observe a decreasing trend of the loss tangent with decreasing electric field with saturation below $\sim10$~V/m, that can be only qualitatively described by the VRH mechanism. More experimental studies are on the way to fully characterize the dissipation of silicon in the milli-Kelvin regime and uncover the detailed nature of the underlying mechanisms at play.

The methodology developed in this study will serve as a tool for detailed investigation of losses in dielectrics at milli-Kelvin temperatures, and guide the selection of materials for the fabrication of high-coherence quantum devices.

This material is based on work supported by the U.S. Department of Energy, Office of Science, National Quantum Information Science Research Centers, Superconducting Quantum Materials and Systems Center (SQMS) under contract number DE-AC02-07CH11359.

\bibliography{Bibliography}
\end{document}